\documentstyle[epsfig,12pt]{article}
\begin{document}
\vskip 1cm
\centerline{\bf FOUR-GAP GLASS RPC AS A CANDIDATE TO}
\centerline{\bf A LARGE AREA THIN TIME-OF-FLIGHT DETECTOR}
\vskip 1cm
\centerline{ Ammosov~V., Gapienko~V.,
Semak~A., Sen'ko~V,}
\centerline{Sviridov~Yu., Zaets~V. Usenko~E.}
\vskip 0.5cm
\centerline{Institute for High Energy Physics, Protvino, Russia}
\vskip 1cm

\begin{abstract}

A four-gap glass RPC with 0.3~mm gap size  was
tested with hadron beam as a time-of-flight detector
having a time resolution of $\sim 100~ps$. A thickness of the detector
together with front-end electronics is  $\sim 12$~mm. 
Results on time resolution dependently on a pad size are 
presented. This paper contains 
first result on the timing RPC (with $\sim 100~ps$ resolution) having 
a strip read-out. Study has been done within
the HARP experiment (CERN-PS214) R\&D work. 
A obtained data can be useful if a design of a large area thin timing detector 
has to be done.
\end{abstract}

\section*{\bf Introduction.}
\par

This study was initiated by necessity to get
thin timing detector for the HARP experiment (CERN-PS214\cite{HARP}).
A multigap RPC operated in avalanche mode at
atmospheric pressure was suggested to use as a Time-Of-Flight 
(TOF) detector in \cite{WILLIAMS}. Then a set of papers
with results obtained during the ALICE TOF R\&D tests were published
(see, for example,\cite{AKINDINOV},\cite{FONTE1},\cite{FONTE2}). 
However, thickness of prototypes 
(chamber itself plus front-end electronics) developed for the ALICE
experiment  is rather few ten cm  what did not
satisfy a HARP request on a TOF system: it thickness should be 
less than 14 mm. A new R\&D work had to be done. 
\par
As the final result of our work, the TOF system covering 
$10~m^2$ and having 368 readout channels has been realized in the HARP
experiment. Its 
construction and basic characteristics were given in\cite{RPC2001}. 
But a lot of interesting results obtained
during the $R\&D$ study are not published yet and here we try to present
them. 
\par
To minimize a number of resistive electrodes and thus to get
thin detector we fixed our attention on a 4-gap RPC with gap thickness of
$0.3~mm$. Operation of this type
of RPC working as TOF detector was described in\cite{FONTE1}. However,
detector developed in \cite{FONTE1} looks like array of 
$32\times 32~mm^2$ independent cells mounted in 
two layers following a chess-board-like
pattern. In our study we tried to get the  TOF detector consisting  
of one chamber added with set of pick-up electrodes.
\par
A tracking system of the HARP detector provides with coordinates 
of point where a particle crosses the TOF RPC what allows to
do not worry about time propagation along the signal electrode
because it can be taken into account. Testing chambers with
different pad size we tried to understand
how the time resolution depends on  a signal electrode area for
to minimize number of channels in the future TOF system.
Furthermore we checked an idea to summarize (with electronics)
signals arriving from several signal pads. First result 
on timing RPC with strip read-out line is presented. 
\par

\section{\bf Experimental setup.}
Data were obtained at CERN in a T10 test area with the 7~GeV/c 
pion beam. 
In tests we used gas line, DAQ system and other
set-up of the ALICE TOF group.   Four trigger
scintillating  counters  in  the  front  and behind  of  our  chambers
provided a selection  of  beam particles  inside of a $1\times  1~cm^2$
spot. Last fact helps us do  not worry about a time jitter coming from
the difference in the propagation time. Start  scintillating counters 
provided  
with $30~ps$ time accuracy giving time mark for precise  measurements.
As working gas the following tetrafluorethane based mixture was used:
$C_{2}H_{2}F_{4}/C_{4}H_{10}/SF_{6}$(90/5/5).

\subsection{\bf RPC construction} 
\begin{figure}[h]
{\epsfxsize14cm\epsfysize6cm\epsffile{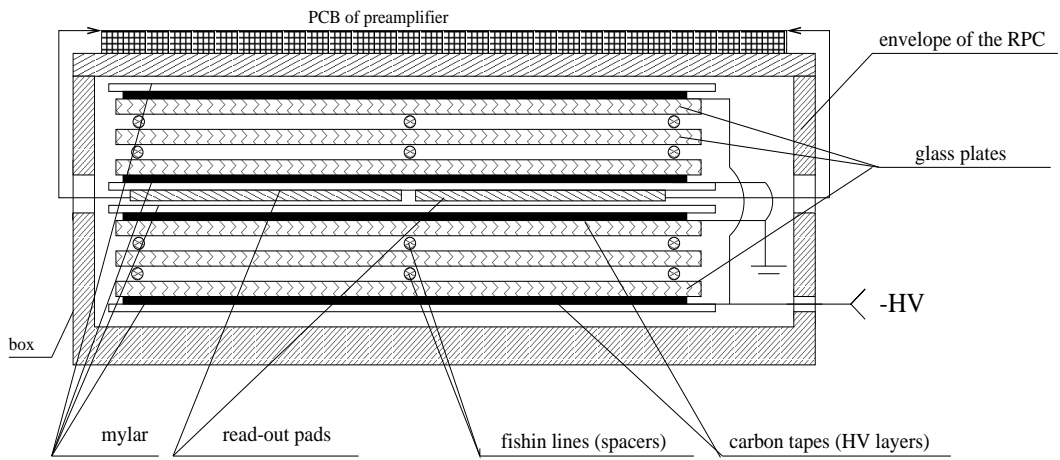}}
\caption{\label{RPC} Construction of the four-gap RPC counter.}
\end{figure}

Several small 4-gap chambers have been built with use of two different
kinds of glass: two chambers having $130\times 200~mm^2$ active area    
were constructed with 1~mm glass taken from the CERN workshop,
one smaller chamber of $70\times 130~mm^2$ size and one long
detector, $70\times 1000~mm^2$, were made of 0.6~mm thick glass
which is used to build prototypes of the ALICE TOF detector.
A resistivity of 0.6~mm glass was measured by us
(with a test voltage of 1~kV) as $\sim 9\times 10^{12}\Omega\star cm$. 
For 1~mm glass plates we found that resistivity is $\sim 7\times 
10^{12}\Omega\star cm$.
A construction  was similar for all  chambers. The cross-section
of four-gap RPC is shown in fig.\ref{RPC}: a pair of
identical double-gap RPCs made of three glass plates 
with 0.3~mm (in diameter)
fishing line as a spacer between plates, and peak-up electrodes,
(pads or strips) between two double-gap chambers.
In the counter made of 0.6~mm glass three fishing lines were put in each
gas gap with spacing of 35~mm between them. Four fishing line
with 40~mm space between them were in chambers made of 
$130\times 200~cm^2$ plates. Few small drops of
'5 minutes' epoxy were enough to fix fishing lines between glass
plates. High voltage was applied  to each of double-gap 
RPCs through electrodes made of high resistive ($\sim 1~M\Omega/\Box$)
carbon film.
Each of five chambers were put in its aluminum box. 
Two $200~\mu m$ mylar
sheets one at the top and one at the bottom provide with an isolation 
between high voltage electrodes and
walls of boxes. Each box can be easily opened and a system of pads
can be changed in few ten minutes.

\subsection{Front-End Electronics}
\begin{figure}
{\epsfxsize12cm\epsfysize8cm\epsffile{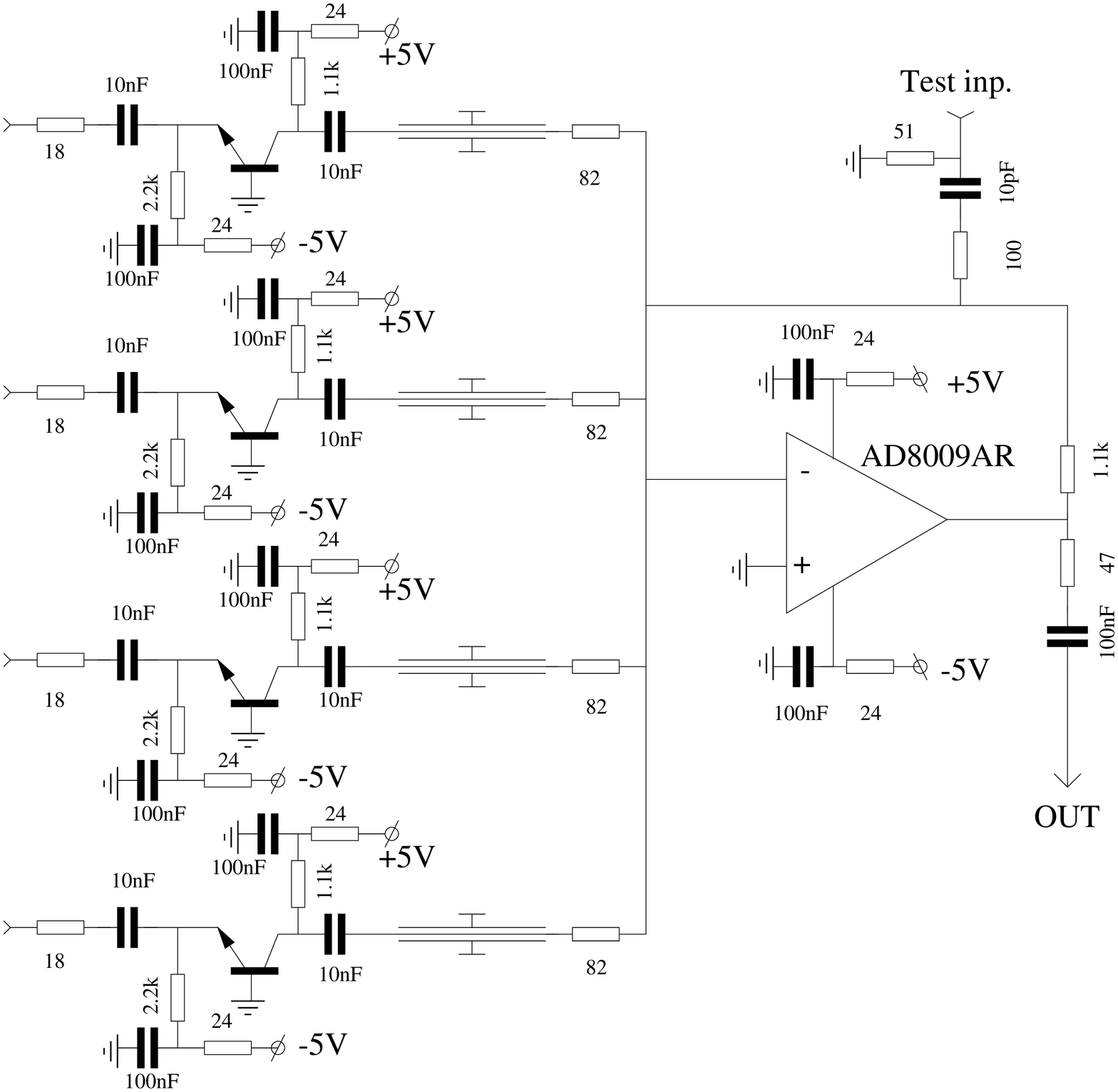}}
\caption{\label{AMP} Scheme of the four-input pre-amplifier.}
{\epsfxsize15cm\epsfysize9cm\epsffile{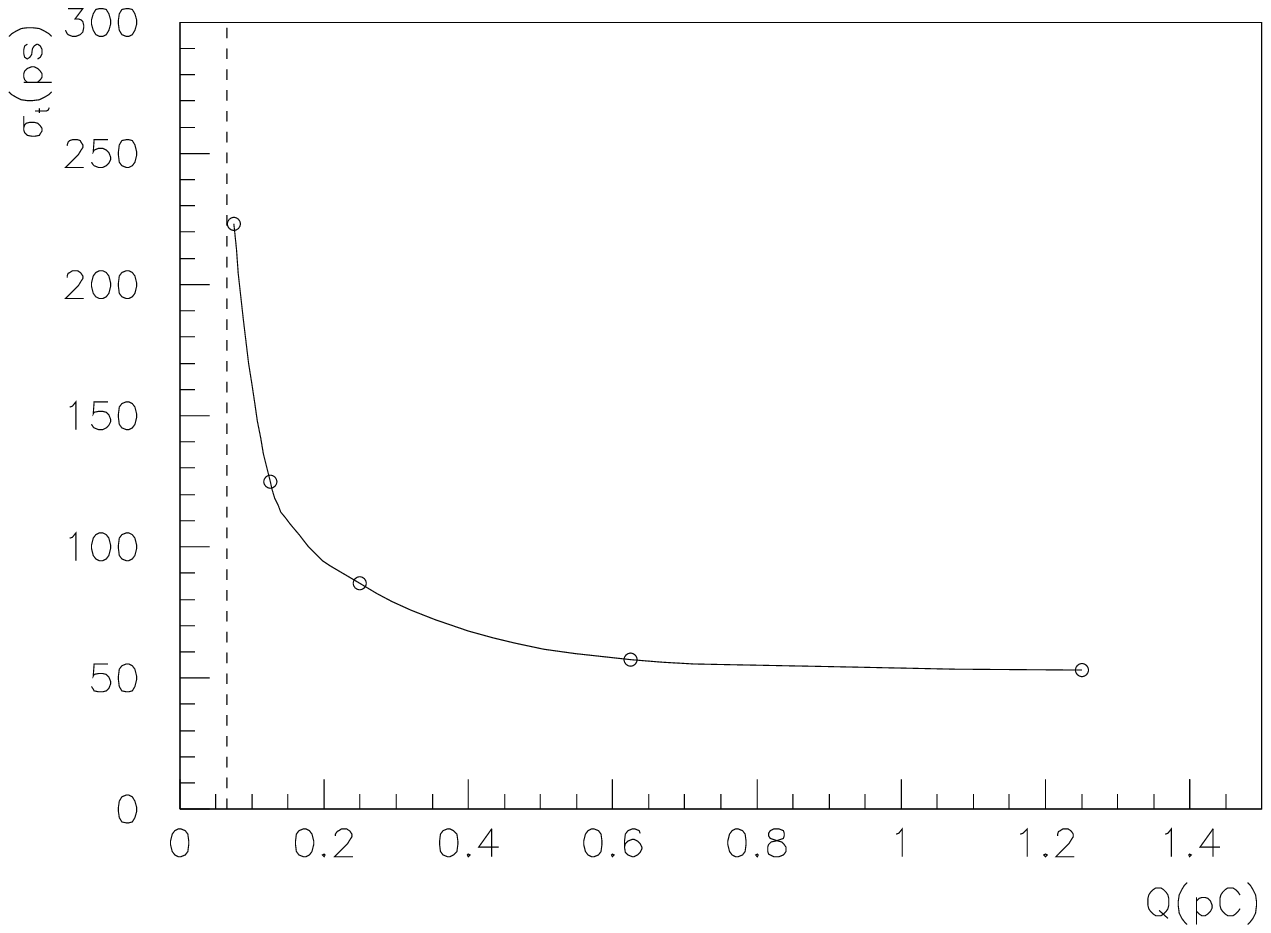}}
\caption{\label{AMP_res} Time resolution of electronics dependently on input charge.}
\end{figure}

A front-end electronics (FEE) used in most of present tests consisted of four-input 
pre-amplifier and splitter/discriminator (SD), both "home-made".
A scheme of the pre-amplifier is presented in fig.\ref{AMP}.
The input circuits contain KT368A9 transistors. Signals from
four inputs are summarized with an AD8009AR amplifier
having 1~GHz bandwidth. A variation
in time of propagation for different inputs is about 10~ps. An
accuracy of the amplitude summarizing is $5~-~7\%$.  
A coefficient of transformation for the pre-amplifier was
measured as -85mV/pC. Inputs of pre-amplifier were connected to
pads with short ($\sim 1-3~cm$) wires. The thickness of 
aluminum box together with pre-amplifier PCB
attached to the RPC cap (as it is schematically shown in fig.\ref{RPC}) 
was about $12~mm$.

\par
A output of the pre-amplifier was connected by the coaxial cable to the  
SD with adjustable ($\sim 3-10~mV$) threshold 
of discrimination.
The SD module has two outputs: one for the analog signal
and one for the discriminated signal (NIM level). The output with analog signal
was directly fed to a LeCroy 2249W ADC. The digital signal was sent to 
a LeCroy 2229 TDC with a $50~ps$ bin width.   
An information from ADC was used to find correlation 
between the "time" and the "amplitude" and then to correct the data on a
time-charge slewing. 
\par
A time resolution of the FEE in described above set-up has been tested with
pulse generator: we injected charge through the test input
of the pre-amplifier and measured the time jitter between the generator pulse 
and output signal from the discriminator as a function of the input charge.
Black circles in fig.\ref{AMP_res} describes the time resolution  
of the pre-amplifier when no one of inputs
was connected to pads.  During this
measurement pre-amplifier was attached to the aluminum box of the
RPC, but no high voltage was applied to the chamber. 
A vertical dashed line in the figure corresponds to chosen by us
threshold of $65~fC$.

\par
Our home-made 4-input pre-amplifier and SD were taken
as a prototypes for the HARP TOF electronics. When first examples of the  
8-input pre-amplifier produced for the HARP detector was available
we used one of them to
see how time resolution changes if to summarize signals
from several pads.

\section{Experimental results.}

\begin{figure}
{\epsfxsize15cm\epsfysize16cm\epsffile{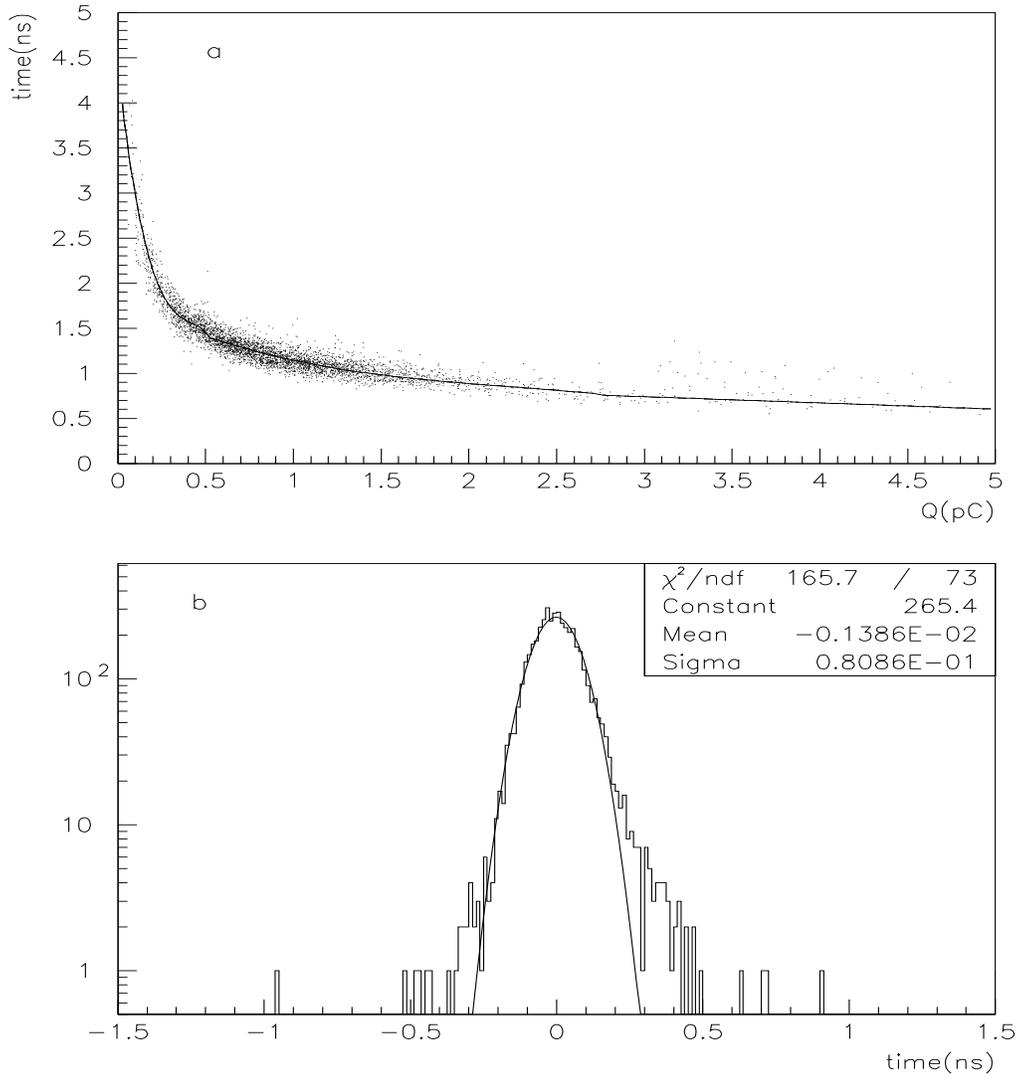}}
\caption{\label{T-Q}~a)-example of charge-time plot: points are
experimental points, curve is the result of approximation
with polynomial expression; b)-time distribution after
correction for the charge-time correlation.} 
\end{figure}

The time resolution of  a detector working near the threshold of electronics
can be estimated  after correction for a time-amplitude correlation. 
A example of typical time-charge scatter plot what we 
saw during  tests of our RPCs is given in fig.\ref{T-Q}a. 
A solid curve in this figure
presents result of the fit   with polynomial
expression. Polynomial functions obtained in such approximation
were used to correct the time distribution for time-charge
correlation. An example of the corrected     
time distribution obtained from the data shown in fig.\ref{T-Q}
is presented in fig.\ref{T-Q}b. Approximation of the corrected
time distribution  with a Gaussian law, as it is shown
in fig.\ref{T-Q}b by curve, gives value of the time resolution ($\sigma_t$).

\subsection{Time resolution and efficiency in dependence on pad size.}

\begin{figure}
{\epsfxsize15cm\epsfysize8cm\epsffile{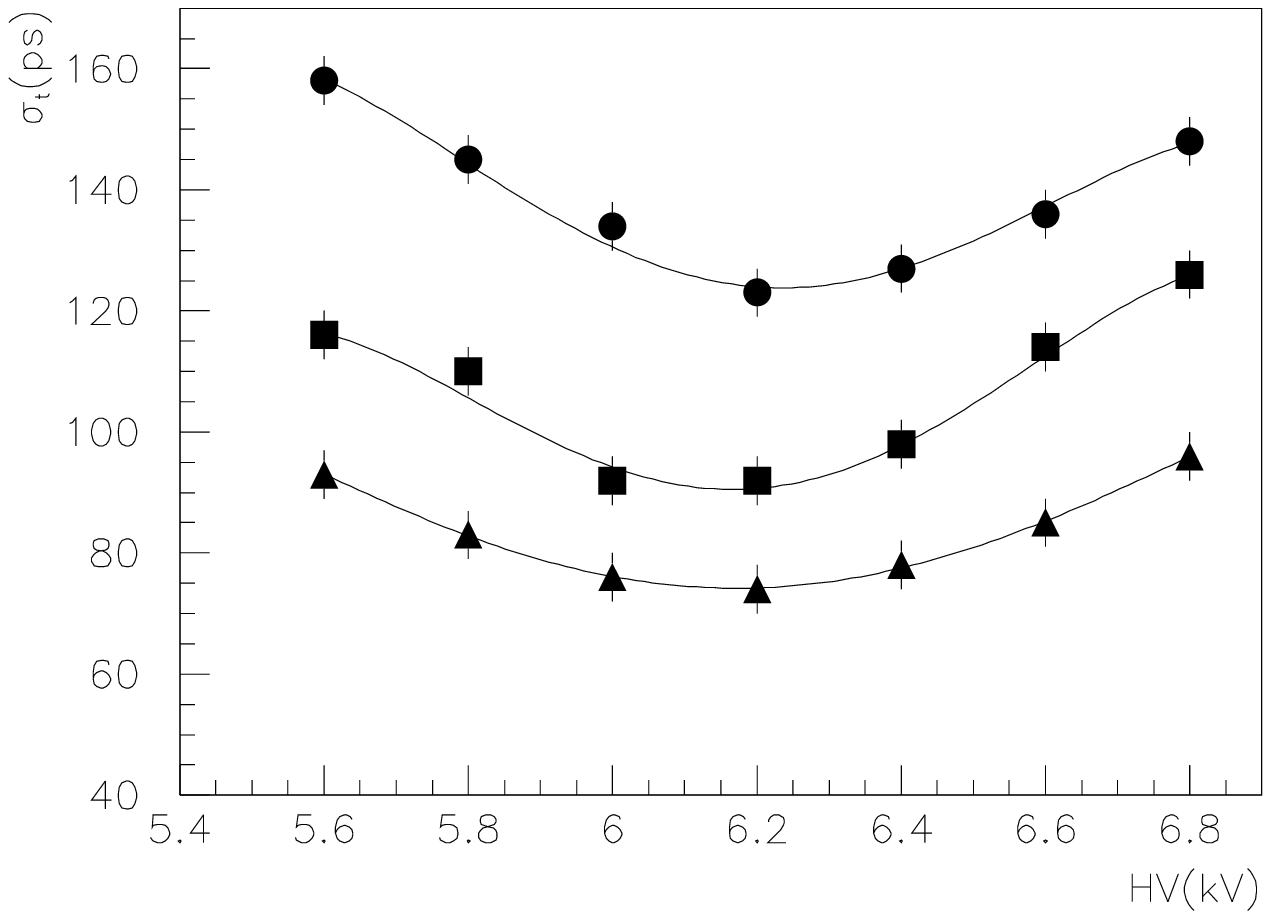}}
\caption{\label{res_HV} Resolution versus high voltage. Triangles, boxes
and circles show data obtained with RPC having 
correspondently $3\times 3~cm^2$, $10\times 10~cm^2$ and $11\times 18~cm^2$
single pad.}
{\epsfxsize15cm\epsfysize8cm\epsffile{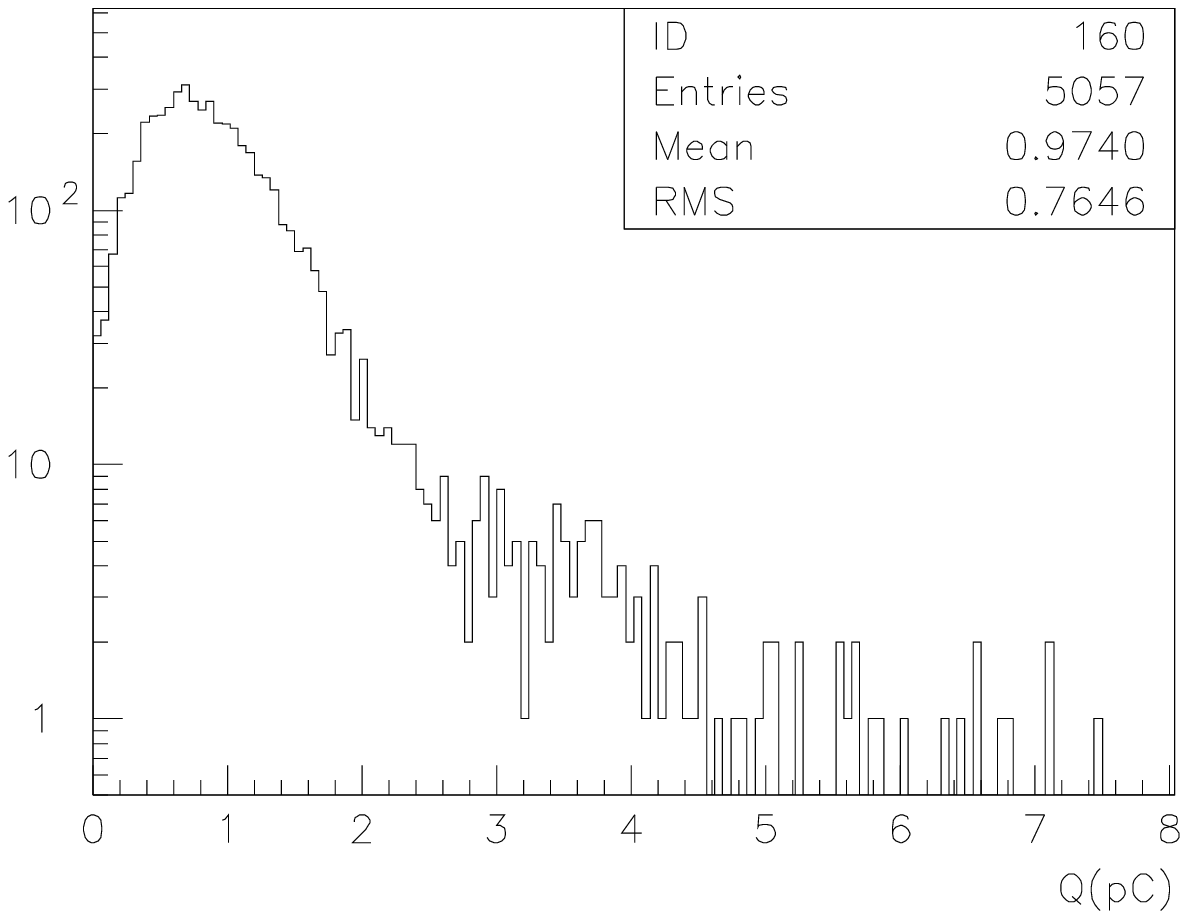}}
\caption{\label{CHARGE} Charge distribution measured at HV=6.2~kV.}
\end{figure}

A working voltage (HV) for the chosen variant of four-gap RPC is
relatively low.  
A behavior of $\sigma_t$ as a function of HV is presented
in fig.\ref{res_HV} as it was observed for chambers with different pad sizes: 
$3\times3$ (triangles),
$10\times 10$ (boxes) and $11\times 18~cm^2$ (circles). Solid curves in the figure are 
drawn for the eyes guide. 
One pad only was connected to the amplifier in last measurement. 
The figure
shows that the best time resolution can be reached at HV=6.2~kV. 
This conclusion was well for all chambers we tested  in spite
of differences in thickness of glass 
used to build chambers.
The value  of 6.2~kV was chosen  as a working voltage in further tests. 
\par
In fig.\ref{res_HV} and in all other figures where data on 
time resolution are plotted,
errors in  $\sigma_t$ shows an accuracy with which we could reproduce our
results next time.
All points presented in fig.\ref{res_HV} were obtained
with chambers passed training under high voltage during several days.
\par
A charge distribution  measured for induced signals
at HV=6.2~kV
is shown in the fig.\ref{CHARGE}. A mean value of  induced charge
is $\sim 1~pC$. Taking into account  the FEE resolution as a
function of the charge (fig.\ref{AMP_res}) and
the real charge distribution  from fig.\ref{CHARGE}, 
one can estimate the  resolution what can be reached in our tests: 
it is $\sim 65~ps$.  Last estimation is ultimate - it does not
includes deterioration in the resolution due to noise
from pads when HV is on. 
\par
Just first our attempts to measure $\sigma_t$ 
showed that the data are unstable with time: improvement
of the resolution was observed  if to keep the chamber
under working voltage for several days. A example is given in fig.\ref{TRAINING} 
demonstrate importance of the training process:
if fresh chamber showed $\sigma_t \approx 170~ps$, after 100 hours the 
resolution drops down below $120~ps$.   
A chamber was considered completely trained, if after several days
of the training process it showed the stable result on $\sigma_t$.

\par
The time resolution dependently on pad area, $S$,  is given 
in fig.\ref{8PADS} with triangles for the case when one pad only was connected
to the pre-amplifier.
An approximation of the data obtained for the single pad 
with expression $\sigma_{t}=A+B\star S$ (solid line in the figure) brings
$A=70~ps$ and $B=0.26~ps/cm^2$. A first value is close to the resolution
what we expected from our electronics for the charge spectrum
shown in fig.\ref{CHARGE}. The  parameter $B$ depends on how noisy chamber is. 
\par
\begin{figure}
{\epsfxsize15cm\epsfysize8cm\epsffile{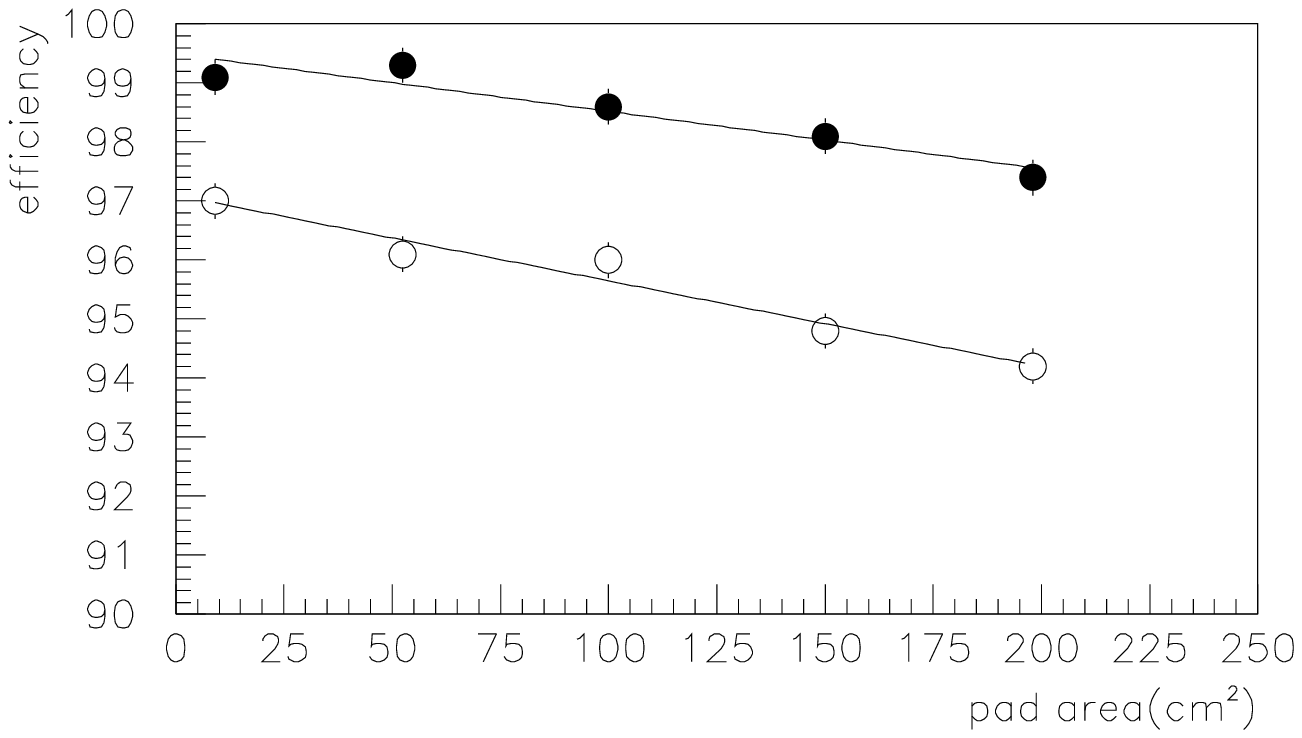}}
\caption{\label{eff_PAD}~Total efficiency (closed circles) and
$3\sigma$-efficiency (open circles) as  functions of pad size.}
{\epsfxsize15cm\epsfysize8cm\epsffile{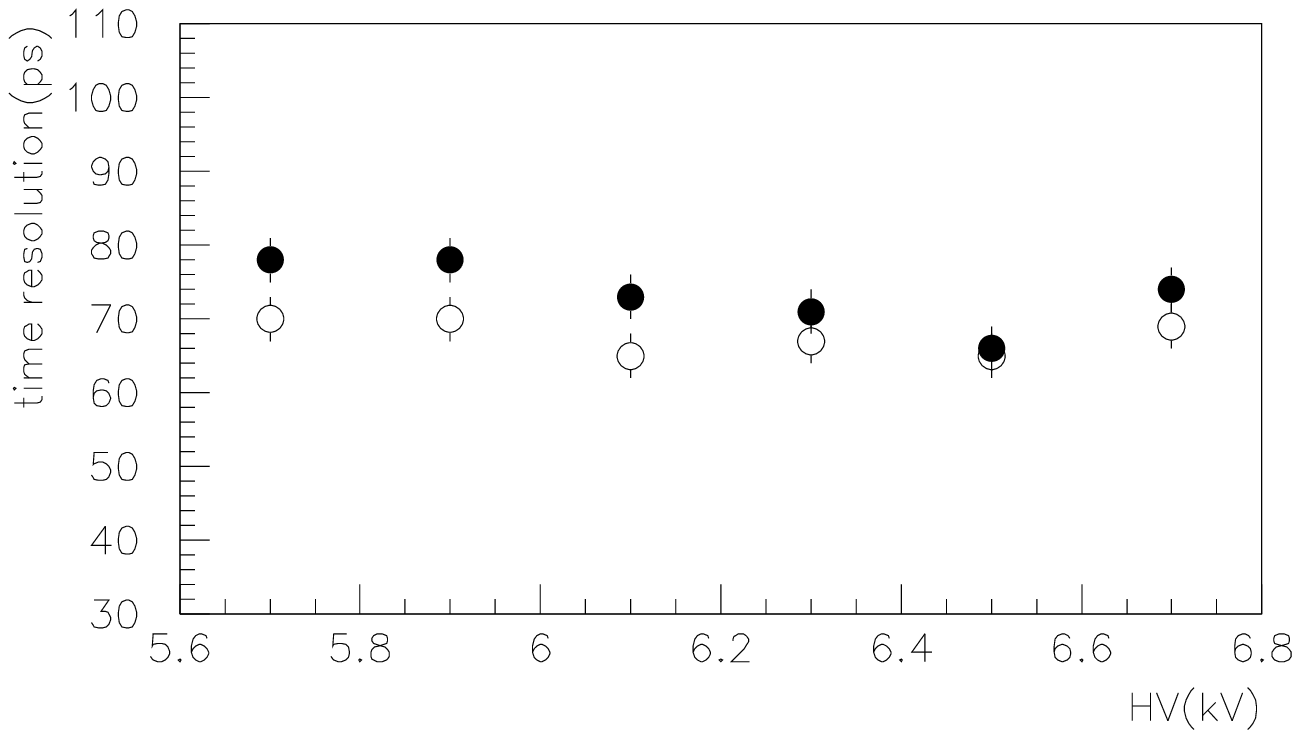}}
\caption{\label{MAX} Corrected time resolution (open circles)
found in case 
when MAX3760 amplifier was followed with standard discriminator  
in comparison with not corrected resolution (closed circles) 
obtained with CFD and two-stage pre-amplifier.}
\end{figure}

All tested RPCs  showed
high efficiency. Two types of efficiency were considered by us:
total efficiency, $\varepsilon$, - probability to get a reply from chamber
inside $15~ns$ gate
and so-called '$3\sigma$-efficiency, $\varepsilon_{3\sigma}$, as a 
probability  
that a reply is inside $\pm 3\sigma_t$ interval. 
Values of $\varepsilon$ (closed circles) and $\varepsilon_{3\sigma}$ 
(open circles) measured for chambers with 
different pads are given in fig.\ref{eff_PAD} as a function of the pad size.
Both efficiencies were calculated, of course, after correction for the
time-charge correlation.
Looking at fig.\ref{eff_PAD} one can see that both efficiencies go down with
growth of the pad size. Equaling to 97\% for $25~cm^2$-pad, the $3\sigma$-efficiency
drops down to $\sim 94\%$ when pad size is $\sim 200~cm^2$.

\par
\begin{figure}
\begin{center}
{\epsfxsize11cm\epsfysize18cm\epsffile{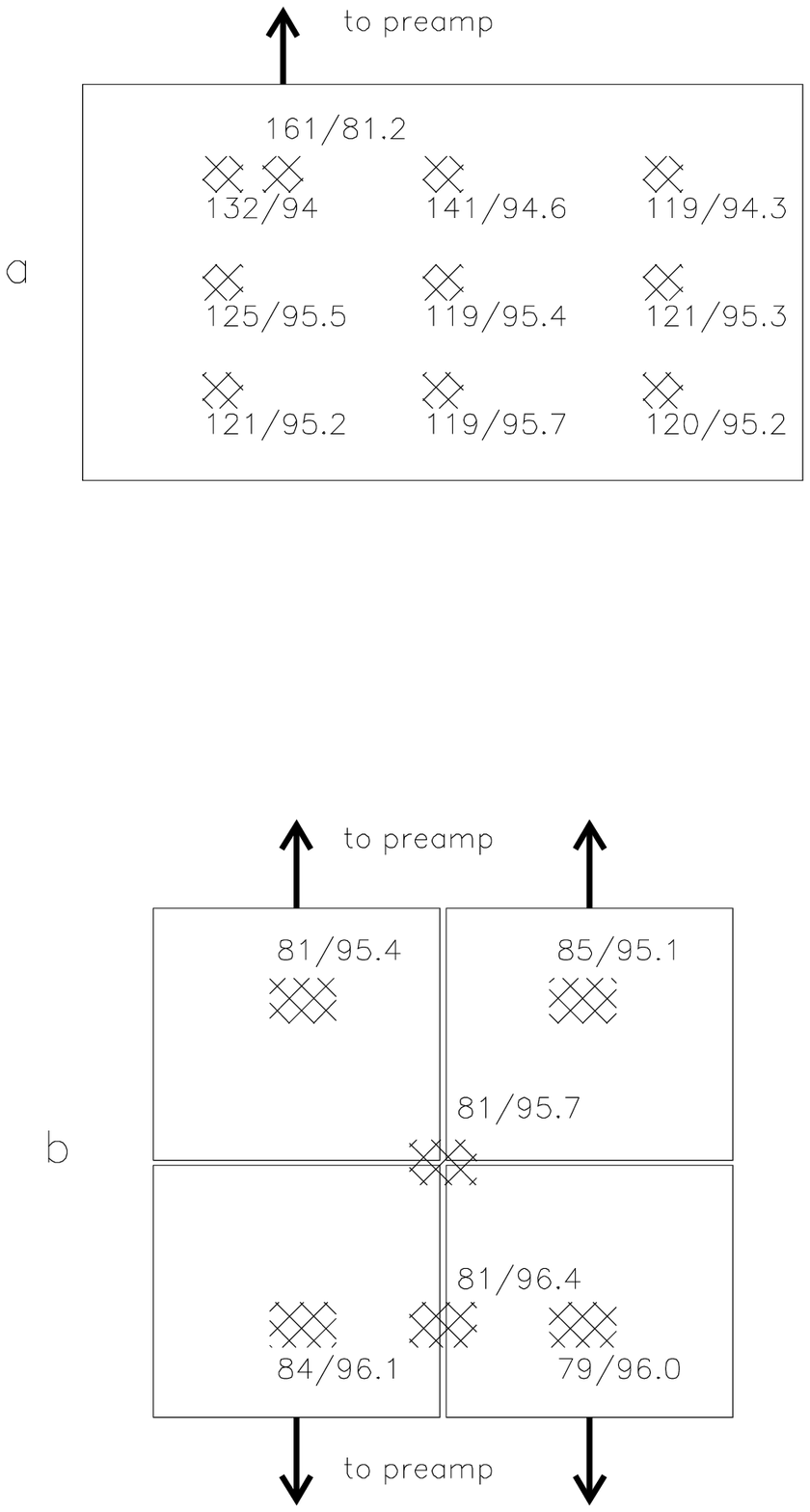}}
\end{center}
\caption{\label{scanpad} Time resolution and efficiency ($\varepsilon_{3\sigma}$)
at different places of a)~$11\times 18~cm^2$ single pad
and b) four-pad system. Size of a pad in last case is $4.3\times 5.6~cm^2$.}
\end{figure}

Another problem was found for big pads.
It was appeared that
for a pad having size of about $10\times 10~cm^2$ and larger 
a value of time resolution is not uniform, it
depends on a point of discharge inside RPC. A example of
such non-uniformity is shown in fig.\ref{scanpad}a for $11\times 18~cm^2$
pad. Ten dashed squares in this figure show places where beam crossed
RPC during the pad scan. Two values at each dashed square 
present the time resolution (first value) and $3\sigma$-efficiency (second value).
A deterioration
in the resolution can be seen in the region close to the point where signal 
wire is soldered
to pad. An attempts to
eliminate  the effect by changing of the amplifier input
impedance brought no positive result.
\par 
All data  in figures where  time resolution is presented dependently 
on pad size were obtained  with beam going through the pad center.
\par
It was described above that  $\sigma_t$ values were
extracted from the time distribution after its off-line 
correction for the time-amplitude  slewing. However, it was
interesting for us to check a possibility to get 'on-line'
timing with use of a constant fraction discriminator (CFD). 
As CFD  we took ORTEC CF4000 module. Because of a high level
of discrimination in the CF4000, a gain of our 'standard' amplifier
was not enough to work with the CF4000.
In operation with the CFD we organized a
chain of two amplifiers: MAX3760 (first stage) and AD8009AR
(second stage, gain=5). Results obtained with two-stage
amplifier are shown in fig.\ref{MAX}. Open circles describes
the resolution obtained when the output of the MAX3760 amplifier was
sent to the our SD. 
The off-line correction for the time-charge correlation was
done before plotting these points. Closed circles
show a data obtained  with the CFD  instead of the usual discriminator.
The figure demonstrates a possibility
to get good time resolution even without off-line correction
if CFD is used.

\subsection{Summing from several pads}

After clear deterioration of the  resolution with growth
of the pad size was observed, 
we tried  to improve situation by  splitting one big pad into
several smaller ones for to read them by one multi-input amplifier.
Firstly we looked does
any non-uniformity  in efficiency and $\sigma_t$ for a system of
small pads connected to one amplifier exist if to expose 
different places of the RPC. To check  this we used the RPC having four
pads with size of $4.3\times 5.6~cm^2$ each. Pads were connected to
different inputs of the 4-input amplifier.
Results of the four-pad system scan are presented in  fig.\ref{scanpad}.
Dashed areas in the figure shows places
where beam crossed the system of four electrodes.
Two values written at each dashed square are: first - time resolution,
second - ${3\sigma}$-efficiency. The figure
demonstrate that in limit of our errors  
no variation in the $\sigma_t$ or $\varepsilon_{3\sigma}$
depending on exposure place  was
observed even in the case when a axis of beam was between
four adjacent pads. A main conclusion from the figure is that the amplifier
summarizes signals without problem for the resolution or efficiency.

\par
\begin{figure}
{\epsfxsize15cm\epsfysize8cm\epsffile{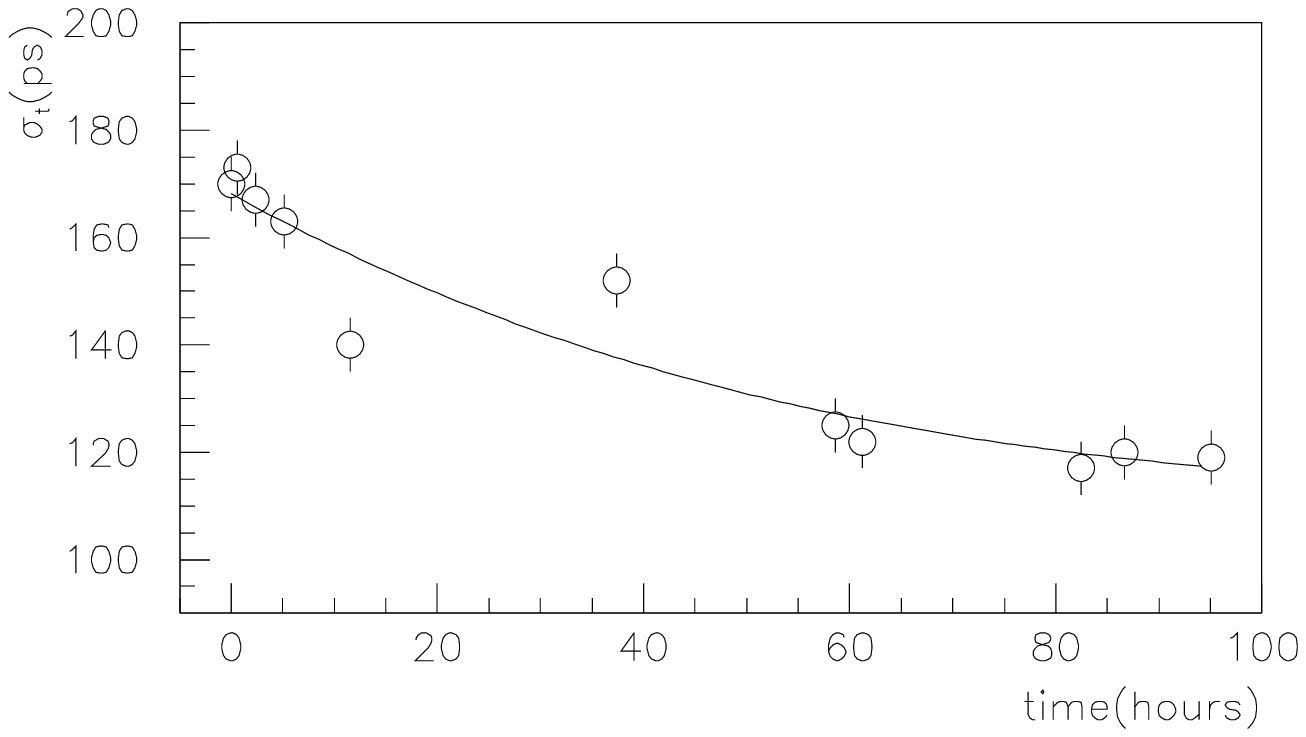}}
\caption{\label{TRAINING} Example of the RPC training: improvement of $\sigma_t$
 with time. Pad area is $11\times 18cm^2$.}
{\epsfxsize15cm\epsfysize8cm\epsffile{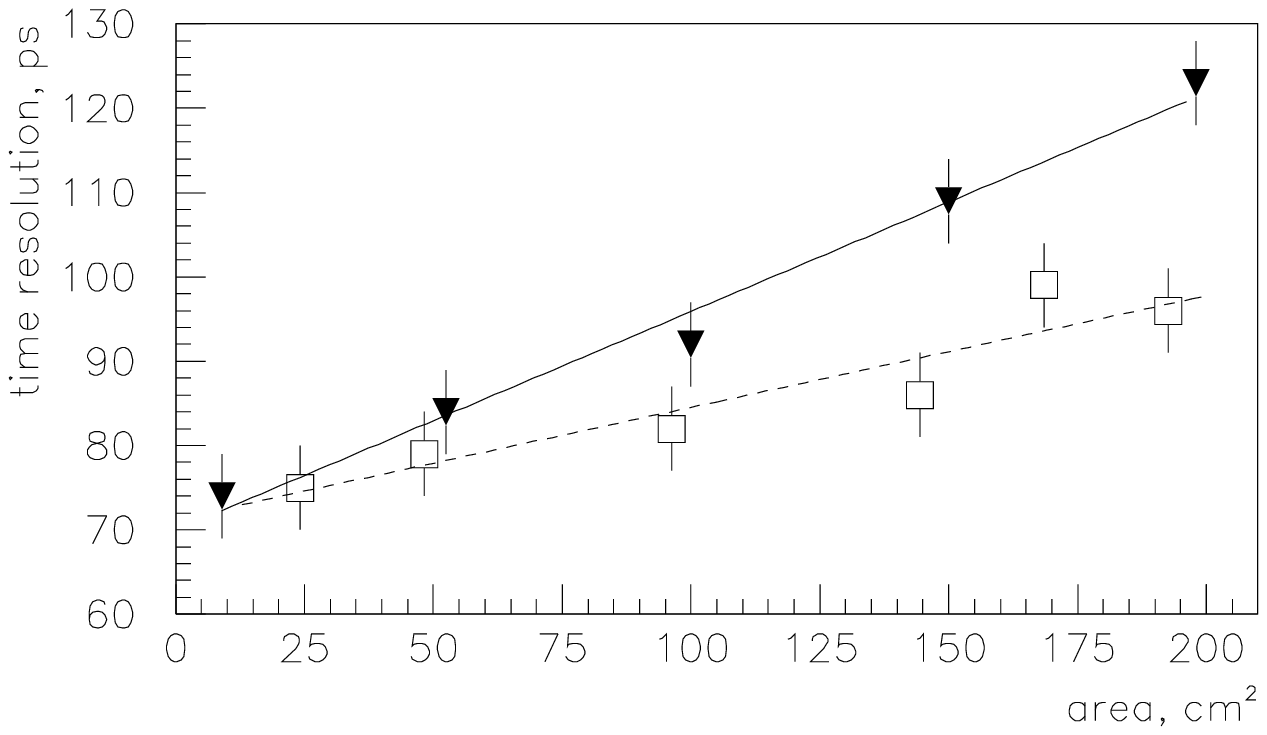}}
\caption{\label{8PADS} Time resolution in case when pre-amplifier reads one
pad (triangles)
and several pads (boxes) as a function of signal electrode area. }
\end{figure}

One module of 8-input
amplifier produced as a final version for the HARP TOF system was 
used by us to see
how resolution depends on 'read-out area' if to summarize
signals from set of small pads. Scheme and main parameters
of final amplifier were practically the same as it was described
above for 4-input. Eight pads  having size  
of $4.3\times 5.6~cm^2$ each were connected (one-by-one)
to eight inputs and the time resolution 
degradation was studied.
Fig.\ref{8PADS} displays $\sigma_t$ as a function of the signal electrodes
area for the case of the single pad (triangles) and for the case when
several pads (boxes)  are connected to one pre-amplifier.
As it is seen from  fig.\ref{8PADS}, a degradation of the resolution
with growth of the area goes
slower for the multi-pad system than for the single pad read-out.
Two lines in the figure presents results of approximation
with a linear law done separately for two sets of the data.
A data obtained with the 8-input amplifier  can
be described with $\sigma_{t}~=~0.71~+~0.13\times S$ (dashed line in
fig.\ref{8PADS}).

\subsection{Strip read-out}

To conclude what the time resolution can be 
reached if a long strip is taken instead of the pad, a 1~m length RPC was
tested with a
$90~cm$ strip line having width of 2.5~cm.
Each end of strip was connected to its own pre-amplifier. 
A impedance of the strip line was measured as $20\pm 5\Omega$.
To avoid possible oscillations due to 
reflection of the signal at ends, the  strip line should be  
terminated with $20\Omega$ loading. 
A resistor was added in the input circuit
of the pre-amplifier to change it impedance to the needed value of $20\Omega$.
\par
\begin{figure}
{\epsfxsize15cm\epsfysize17cm\epsffile{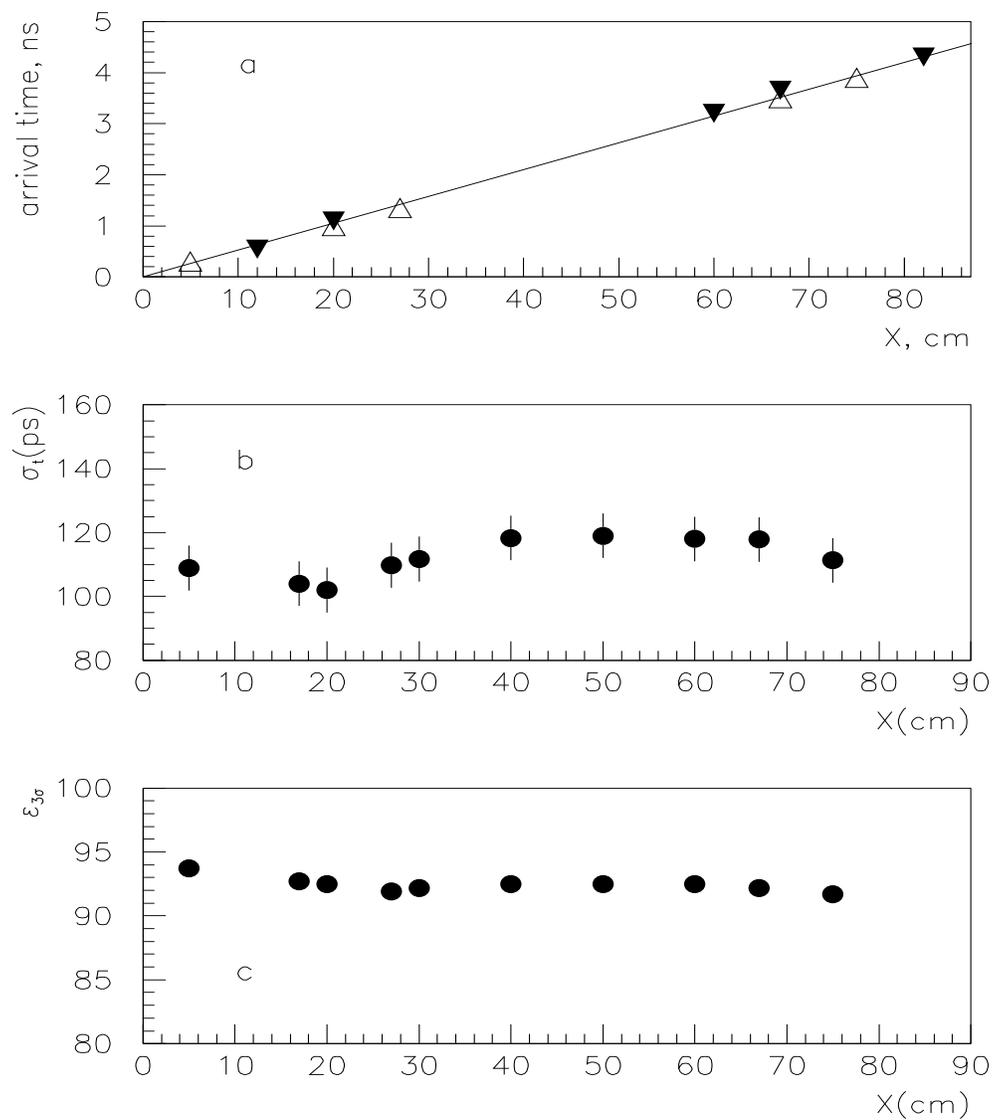}}
\caption{\label{doska} a)- arrival time,  
b)-time resolution and c)-$3\sigma$-efficiency dependently on
distance between strip end and beam.}
\end{figure}

Different places of the  1~m length RPC were exposed to the  beam, 
and a reply of the detector was studied dependently on a distance, $X$,
between the beam line and the strip end.  
Fig.\ref{doska}a shows a mean arrival time as a function of the
$X$. Two sets of points
in the figure, open and closed triangles, belong to different 
ends of one  strip line.
A linear approximation of the data gives a slope of $\sim 50~ps/cm$.
\par
The time jitter for half of a
sum of corrected times coming from opposite ends, $t_{12}=(t_{1}+t_{2})/2$,
was measured along the strip for different distances taken relatively  
to one of ends. The results on $t_{12}$ resolution are 
in fig.\ref{doska}b.
As the figure demonstrates,
values of the resolution  found during "scanning" along the strip are 
inside of 110-120~ps. In the limit of experimental errors they do not depend
on the $X$. If to read one end of the strip only, the time resolution was
found to be about $160~ps$. 
\par
Fig.\ref{doska}c presents the ${3\sigma}$-efficiency estimated
from a  width of $t_{12}$-distribution. 
No dependence  of $\varepsilon_{3\sigma}$ on $X$ is seen in the last figure.

\subsection{Glass RPC at different particle rate}
\par
The glass is a high resistive material. That is why the glass RPC
even operated in the avalanche mode should be  sensitive
to a  particle rate. 
\par
Within set-up  what we had in the T10 test area
we estimated the rate using 
counting rates from set of scintillating counters
installed in front and behind of our chambers.
However, we could not control particle rate with accuracy
better than $\sim 30\%$, because 
few detectors tested by other groups were installed at the
same time in the beam line.
\par
The rate what we show in the figures below
was calculated as a number of particle crossing  
$10\times 10~mm^2$ square (with center at beam line)
per spill divided by spill duration of $\sim 300~ms$. 
It should be noticed that
because of a beam halo not only $1\times 1~cm^2$ RPC area  was
exposed to the beam.
A number of particles going through  a $4.5\times 5~cm^2$ scintillating
counter installed just before tested RPC was 
$\sim 15$ times higher than the
number of particles going trough the central $10\times 10~mm^2$ area.
\par
\begin{figure}
{\epsfxsize15cm\epsfysize17cm\epsffile{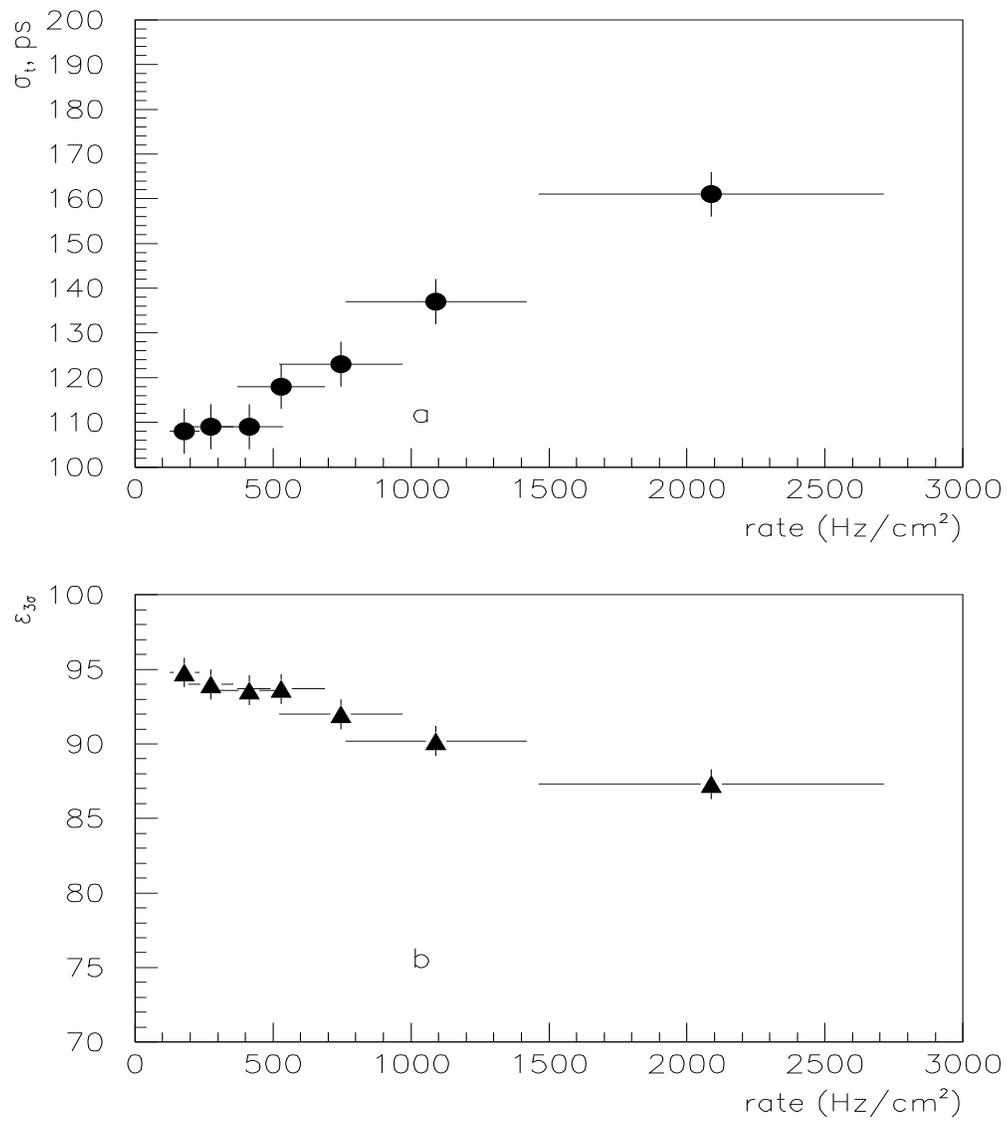}}
\caption{\label{RATE} a- time resolution vs particle rate,
b- efficiency at different rates}
\end{figure}

Fig.\ref{RATE} shows: (a) time resolution  and (b) $\varepsilon_{3\sigma}$
efficiency at different values of the  rate. 
The RPC having $10\times 15$ single pad was exposed to beam
to get these distributions. Both distributions shown in fig.\ref{RATE}
demonstrate a deterioration in the RPC work with growth of the particle rate.

\section*{Conclusions}
\par
We designed, built and tested prototypes of the thin TOF detector.
Four-gap glass RPCs added with different pick-up electrodes
were exposed to $7~Gev/c$ pion beam to find what the time resolution and registration efficiency
can be reached. 
\par
The time resolution as a function of the pad size show
linear dependence on pad area ($S$): $\sigma_t$=$A~+~B\times S$,
where $A~=~70~ps$ is the intrinsic resolution of our FEE, and
$B~=~0.26 ps/cm^2$.  
The value of $3\sigma$-efficiency is 97\% for the small pad ($25~cm^2$)
and drops down to 94\% with growth of the pad area to $\sim 200~cm^2$. 
\par
The situation with the time resolution and efficiency 
for the 'big' read-out area can be improved if
to split one pad into set of smaller ones and to read them 
with  summing amplifier.
Using  8-input amplifier we found that 
$\sigma_{t}~=~0.71~+~0.13\times S$. 
\par 
We tested the timing RPC with the strip read-out. The resolution of $110-120~ps$
was reached for the 90~cm length strip having 2.5~cm width when
signals from both ends were recorded.
\par
Our results can be useful if development of large area
TOF detector with thickness of $10-15~mm$ is needed.
\vskip 0.3cm
\par
{\bf Acknowledgment}
We would like to thank members of the ALICE TOF group, especially 
M.C.S Williams for enormous help with materials and
equipment. 
\par
Furthermore we are very grateful to F.Dydak and J.Wotschack
for their  suggestions and for their constant
interest on our study.

\vskip 2cm

\end{document}